\documentclass[journal,draftcls,onecolumn,12pt,twoside]{IEEEtranTCOM}

\normalsize

\usepackage{amsmath}
\usepackage[dvips]{graphicx}
\usepackage{amssymb}
\usepackage{array}
\usepackage{multirow}

\begin{document}
%
\title{TDCS-based Cognitive Radio Networks with Multiuser Interference Avoidance}

\author{Su~Hu,~\IEEEmembership{Member},
        Guoan~Bi,~\IEEEmembership{Senior Member},
        Yong~Liang~Guan,~\IEEEmembership{Member},
        and~Shaoqian~Li,~\IEEEmembership{Senior Member}
\thanks{S. Hu and S. Li are with National Key Laboratory of Science and Technology on Communications, University of Electronic Science and Technology of China (UESTC), China. e-mail: (husu@uestc.edu.cn).}
\thanks{G. Bi and Y.L. Guan are with School of Electrical and Electronic Engineering, Nanyang Technological University (NTU), Singapore.}}

\markboth{IEEE Transactions on Communications}%
{Submitted paper}
\maketitle

\begin{abstract}
For overlay cognitive radio networks (CRNs), transform domain communication system (TDCS) has been proposed to support multiuser communications through spectrum bin nulling and frequency domain spreading. In TDCS-based CRNs, each user is assigned a specific pseudorandom spreading sequence. However, the existence of multiuser interference (MUI) has been seen one of main concerns, due to the non-zero crosscorrelations between any pair of TDCS signals. In this paper, a novel framework of TDCS-based CRNs with the joint design of sequences and modulation schemes is presented to realize MUI avoidance. With the uncertainty of spectrum sensing results in CRNs, we first introduce a sequence design through two-dimensional time-frequency synthesis and obtain a class of almost perfect sequences. That is, periodic autocorrelation and crosscorrelations are identically zero for most circular shifts. These correlation properties are further exploited in conjunction with a specially-designed cyclic code shift keying in order to achieve the advantage of MUI avoidance. Numerical results demonstrate that the MUI-free TDCS-based CRNs are considered as preferable candidates for decentralized CRNs against the near-far problem.
\end{abstract}

\begin{IEEEkeywords}
Cognitive radio network, transform domain communication system, multiuser interference, periodic correlation function.
\end{IEEEkeywords}

\section{Introduction}
\IEEEPARstart{D}{ue} to the scarcity of available spectrum, cognitive radio (CR) is considered to be a promising paradigm to provide the capability of using or sharing  the spectrum in an opportunistic manner \cite{Haykin}. Multicarrier modulations are preferable candidates for realizing CR applications in which arbitrarily selected subcarriers are disabled to avoid using the occupied spectrum bands \cite{Boroujeny}. Following a general analytic framework that is spectrally modulated and spectrally encoded \cite {RobertSP}\cite{Chakrava}, various multicarrier waveforms can be generated based on CR user needs, e.g., orthogonal frequency division multiplex (OFDM) and multicarrier code division multiple access (MC-CDMA) \cite{Hara}. As another type of overlay CR, transform domain communication system (TDCS) has been proposed to support reliable communications with low spectral density through spectrum bin nulling and frequency domain spreading \cite{Budiarjo}-\cite{Cluster}. Hence, the TDCS becomes attractive as a smart anti-jammer transmission scheme when some available spectrum bands are occupied by intentional jammers \cite{Chakravarthy-TDCS}.

To achieve multiple access capability, CR networks (CRNs) share all available resources by the different protocols, e.g., time division multiple access or orthogonal frequency division multiple access \cite{Letaief}. If the code dimension is interpreted as a part of the spectrum space, new opportunities for spectrum usage can be created at a given time based on spectrum spreading \cite{Tevfik}.
Since secondary users in CRNs have to identify the presence of primary users and vacate the occupied spectrum bands immediately whenever primary users are active, the spectrum opportunity in this manner naturally results in the uncertainty of spectrum utilization pattern. Under such CR constraints, those existing spreading schemes, e.g., MC-CDMA, are with random-spreading. Importantly, the (quasi-)orthogonality of spreading codes, e.g., Walsh-Hadamard \cite{PoPo}, and Zadoff-Chu \cite{Zadoff}\cite{Chu}, is destroyed \cite{Feng}. In contrast, when the entire spectrum band is available, multiple users in those schemes are assigned unique and orthogonal spreading codes.

Similarly in traditional TDCS-based CRNs, when each user is assigned a specific pseudorandom polyphase sequence \cite{Swackhammer}, multiuser interference (MUI) becomes a main factor of bit error rate (BER) performance degradation due to the non-zero crosscorrelations between any pair of TDCS signals. This is especially true in decentralized wireless networks in which the phenomenon known as near-far effect has been seen to disrupt communications.

With the uncertainty of spectrum utilization pattern in CRNs, a novel framework of TDCS-based CRNs is presented in this paper to achieve MUI avoidance from the viewpoint of sequence design. Instead of only frequency spreading used in traditional schemes, we first introduce a sequence design through two-dimensional time-frequency synthesis. It is shown that if two short sequences are carefully chosen, a class of almost perfect Kronecker sequences is obtained. That is, periodic autocorrelation and crosscorrelation functions (ACF/CCF) are identically zero for most circular shifts. Importantly, such correlation properties are always guaranteed for any spectrum utilization pattern in CRNs. In addition, since the periodic CCF between any pair of TDCS signals is highly dependent on cyclic code shift keying (CCSK), sufficient attention should be paid for the joint design of almost perfect sequences and modulation schemes. Accordingly, a specially-designed CCSK over a limited shift range is addressed in order to guarantee the mutual orthogonality of TDCS signals for the advantage of MUI avoidance.

This paper is organized as follows. Section II reviews TDCS multiple access applications in order to briefly explain the disadvantages of traditional TDCS-based CRNs. Section III first presents the sequence design through two-dimensional time-frequency synthesis. Then, a specially-designed CCSK is addressed to guarantee the mutual orthogonality of TDCS signals for the advantage of MUI avoidance. In Section IV, multiuser capability and aggregated throughput are discussed respectively. Finally, simulation results are presented in Section V.

\section{Overview of TDCS Multiple Access}
In CR systems, the entire spectrum band is divided into $N$ spectrum bins and a spectrum marking vector, $ {\mathbf{{S}}} = \left\{ {S(0), S(1), \cdots, S(N - 1)} \right\}$, is used to indicate the status of spectrum bins. For example, the value of $S(k)$ is set to 1 (or 0) if the spectrum sensing output for the $k$-th bin is smaller (or larger) than a given threshold. We assume the set of $N_C$ available spectrum bins as $\Omega ^C$, i.e., $\left\{ {S(k) = 1, k \in \Omega ^C } \right\}$, and the same spectrum marking vector is obtained by the spectrum sensing facilities on the receiver side.

On the TDCS transmitter side, a spectral vector, $\mathbf{{B}}$, is obtained by the operation of element-by-element multiplication, i.e., $\mathbf{{B}}=\mathbf{{S}} \cdot diag\left(\mathbf{{P}} \right)$,  where $ {\mathbf{P}} = \left\{ e^{jm(0) }, e^{jm(1) }, \cdots, e^{jm({N - 1}) } \right\}$ is a pseudorandom polyphase sequence, and $m({k}), k\in\{0,1,\cdots,N-1\},$ is the phase shift for the $k$-th bin. Then, a fundamental modulation waveform (FMW), ${\mathbf{{b}}}= \left\{{b(0), b(1), \cdots, b(N-1)} \right\}$, is obtained by performing an $N$-point inverse fast Fourier transform (IFFT) operation on the spectral vector, ${\bf{{B}}}$, \cite{Chakravarthy-TDCS}
\begin{equation}
\label{b-FMW}
b(n)  =  \lambda \sum \limits_{k = 0}^{N-1} {{S(k) {e^{j{m(k)}}} e^{\frac{j 2 \pi k n}{N}}}} = \lambda \sum \limits_{k \in \Omega^C} {e^{j{m(k)}}e^{\frac{j2\pi k n}{N}}},
\end{equation}
where $\lambda = \sqrt {{N / N_C }}$ is a normalization for the transmitted energy. When $M$-ary CCSK is adopted, the transmitted signal, ${\mathbf{{x}}}$, is obtained by shifting ${\mathbf{{b}}}$ cyclically to the left by $\tau \in \left\{0, 1, \cdots, M-1 \right\}$ places,
\begin{equation}\label{Trad}
{\mathbf{{x}}} = \left\{x(0), x(1), \cdots, x(N-1) \right\} = \langle {\mathbf{{b}}} \rangle_{\tau},
\end{equation}
where $\left\langle \cdot \right\rangle_N$ denote $N$-point cyclic shift.

On the TDCS receiver side, the received signal, ${\mathbf{{r}}} = \left\{ r(0), r(1), \cdots, r(N-1)\right\}$, is periodically correlated with the replica of FMW to recover data symbols by detecting the correlation peak, which can be performed by using the fast Fourier transform (FFT) to minimize the computational complexity \cite{Dillard}. Since the information on $\tau$ is hidden in the background noise by spreading over all available spectrum bins, TDCS signals have a low probability of being intercepted by unauthorized listeners. More detailed information on the transmission and reception of TDCSs is available in \cite{Chakravarthy-TDCS}.

Let us consider a $U$-user traditional TDCS-based CRN in which each user is assigned a specific pseudorandom polyphase sequence to share available spectrum resources. A class of polyphase sequences is given by \cite{Swackhammer}
\begin{eqnarray}
\label{PSET}
{\bf{P}} & = & \left\{ {{{\bf{P}}_1}, {{\bf{P}}_2}, \cdots, {{\bf{P}}_{U}}} \right\} \nonumber \\
{{\bf{P}}_i} & = & \left\{ e^{jm_{i}(0) }, \cdots, e^{jm_{i}(k) }, \cdots, e^{jm_{i}(N-1) } \right\},
\end{eqnarray}
where ${{\bf{P}}_i}$ denotes the pseudorandom sequence for the $i$-th user and $e^{jm_{i}(k) }$ denotes the $k$-th sequence element of ${{\bf{P}}_i}$. Based on \eqref{b-FMW}, a class of FMWs assigned to multiple users is given by
\begin{eqnarray}\label{b-array}
{\bf{b}} & = & \left\{ {{{\bf{b}}_1}, {{\bf{b}}_2}, \cdots, {{\bf{b}}_{U}}} \right\} \nonumber \\
{{\bf{b}}_i} & = & \left\{ {b_i(0), b_i(1), \cdots, b_i(N - 1)} \right\} \nonumber \\
b_i(n) & = & \lambda \sum \limits_{k \in \Omega^C} {e^{j{m_{i}(k)}}e^{\frac{j2\pi k n}{N}}}.
\end{eqnarray}
From the previously reported results in \cite{Swackhammer}, we have the periodic CCF between any pair of FMWs $\left( i, j \right)$ as follows,
\begin{equation}
\label{PCCFORI}
{\varphi_{{{\bf{b}}_i},{{\bf{b}}_j}}}\left( \tau  \right) = \sum\limits_{n = 0}^{N - 1} {b_i(n)} {b_j^*({\mathrm{mod}(n + \tau ,N) } )} \neq 0, \forall \tau,
\end{equation}
where $\left( \cdot \right)^*$ denotes the complex conjugate and $\mathrm{mod} \left( x, y \right)$ denotes the remainder of $x$ divided by $y$.

Observed from \eqref{PCCFORI}, it is clear that the MUI always exists in traditional TDCS-based CRNs due to the non-zero periodic crosscorrelations. Especially, the phenomenon known as near-far effect in decentralized wireless networks, such as CR Ad-hoc wireless sensor networks, further aggravates the MUI, leading to  significant BER performance degradation. Therefore, the aim of this paper is to discuss how to design TDCS-based CRNs with MUI avoidance.
	
\section{TDCS-based CRNs with MUI Avoidance}
In this paper, a novel framework of TDCS-based CRNs with the joint design of sequences and modulation schemes is presented to realize MUI avoidance. With the uncertainty of spectrum sensing results in CRNs, we first introduce a sequence design through two-dimensional time-frequency synthesis and obtain a class of almost perfect sequences. Then, a specially-designed modulation scheme is discussed in order to achieve the advantage of MUI avoidance.
\subsection{Design of almost perfect FMWs}
Let ${{\bf{A}}} = \left\{ {a(0), a(1), \cdots, a(L - 1)} \right\}$  be a polyphase sequence of length $L$ having perfect periodic ACF as follows, \cite{Zadoff}
\begin{equation}
\label{PerfectAC}
{\varphi_{{{\bf{A}}}}}\left( \tau  \right) = \sum\limits_{l = 0}^{L - 1} {a(l)} {a^*({\mathrm{mod} \left( {l + \tau}, L \right)}})  = \left\{ {\begin{array}{*{20}{c}}
L, &{\left( {\tau  = 0} \right)}\\
0, &{\left( {\tau  \ne 0} \right)}
\end{array}} \right..
\end{equation}
By taking the sequence, ${{\bf{A}}}$, and the basis FMWs, ${{\bf{b}}}$, from \eqref{b-array}, we obtain a class of Kronecker sequences, ${\bf{c}}$, constructed through two-dimensional time-frequency synthesis as shown in Fig.~\ref{FIG:BLKTX},
\begin{eqnarray}
\label{Krongen}
{\bf{c}} & = & \left\{ {{{\bf{c}}_1},{{\bf{c}}_2}, \cdots, {{\bf{c}}_{U}}} \right\} \nonumber \\
{\bf{c}}_i & =& \left\{c_i(0), c_i(1), \cdots, c_i(LN-1) \right\} \nonumber \\
c_i(n)& = & a\left(l\right) b_i\left(m\right),
\end{eqnarray}
for $n=lN+m, 0 \leq l \leq L-1, 0 \leq m \leq N-1$. According to the correlation property of Kronecker sequences in \cite{Stark}, the periodic CCF $\varphi_{{\bf{c}}_i, {\bf{c}}_j}\left(\tau\right)$ between two sequences ${{\bf{c}}_i}$ and ${{\bf{c}}_j}$ is given by
\begin{equation}
\label{CCKron}
\varphi_{{\bf{c}}_i, {\bf{c}}_j}\left(\tau\right) = \varphi_{{\bf{A}}}\left(l\right) \psi_{{{\bf{b}}_i},{{\bf{b}}_j}}\left( m \right) + \varphi_{{\bf{A}}}\left( l+1 \right) \psi_{{{\bf{b}}_i},{{\bf{b}}_j}}\left( m-N \right),
\end{equation}
where $\tau=lN+m, 0 \leq l \leq L-1, 0 \leq m \leq N-1$, and $\psi _{{\bf{u}},{\bf{v}}}\left( \cdot \right)$ denotes the aperiodic CCF between two sequences $\bf{u}$ and $\bf{v}$ of length $N$ as
\begin{eqnarray}
{\psi _{\bf{u,v}}}\left( \tau \right) = {\sum\limits_{i = 0}^{N - 1 - \tau} {u\left( i \right){{{v^*\left( {i + \tau} \right)} }}},} & {0 \le \tau < N} \nonumber.
\end{eqnarray}
Substituting \eqref{PerfectAC} into \eqref{CCKron}, we obtain the periodic CCF between two designed sequences ${{\bf{c}}_i}$ and ${{\bf{c}}_j}$ as follows,
\begin{equation}
\label{PCCFFMW}
{\varphi _{{{\bf{c}}_i},{{\bf{c}}_j}}}\left( \tau  \right) = \left\{ {\begin{array}{*{20}{l}}
{L{\psi _{{{\bf{b}}_i},{{\bf{b}}_j}}}\left( \tau  \right),}&{ - N < \tau  < N}\\
{0,}& {\left| \tau \right| \ge N}
\end{array}} \right..
\end{equation}
Note that, the periodic ACF of the $i$-th designed sequence is $\varphi_{{{\bf{c}}_i},{{\bf{c}}_i}}\left( \tau  \right)$.

\begin{figure}
\centering
  \includegraphics[width=4.5in]{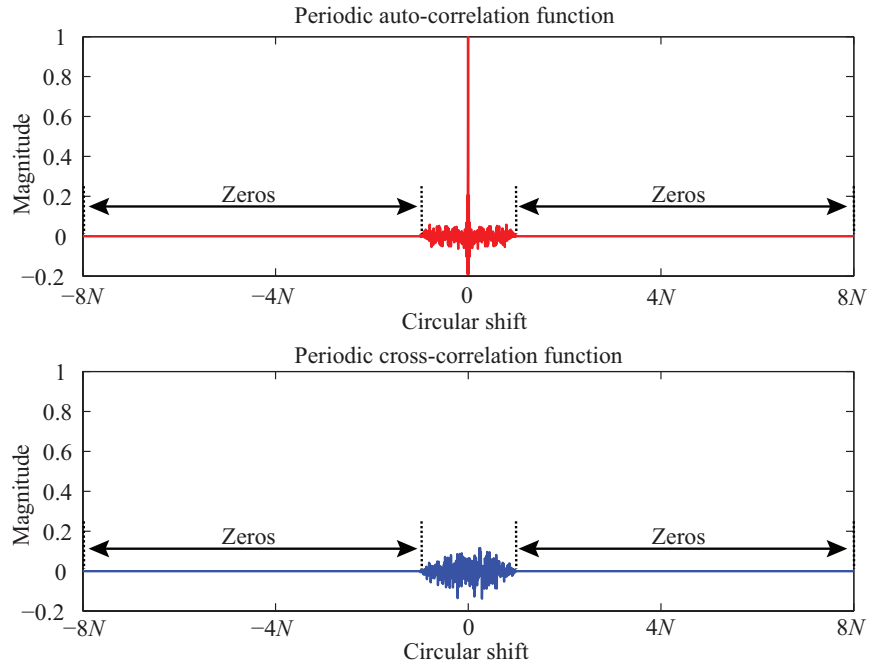}
  \caption{Periodic ACF/CCF of the designed almost perfect sequences ($N$=64, $L$=16)}
  \label{FIG:PCFMUI}
\end{figure}

One particularly valuable property of the designed sequences is that \emph{the periodic auto/cross correlation functions are identically zero for $\left(L-2 \right)N+1$ different values of $\tau$}, as shown in Fig.~\ref{FIG:PCFMUI}. Importantly, the correlation property is independent on the basis FMWs. In other words, it is always guaranteed for any spectrum utilization pattern in CRNs. In addition, even for $|\tau| < N$, both periodic ACF/CCF sidelobes are trivial compared to the periodic autocorrelation mainlobe. Consequently, it is reasonable to consider the designed Kronecker sequences as a class of almost perfect sequences under CR constraints. In the following contents, let us discuss how to utilize the designed sequences in TDCS-based CRNs to achieve the advantage of MUI avoidance. Especially, two scenarios of single-path and multipath fading channels are discussed, respectively.

\subsection{Design of modulation schemes in single-path fading channels}
For $U$-user synchronous TDCS-based CRNs, each user chooses one of the designed sequences in \eqref{Krongen} as the FMW. When $M$-ary CCSK is adopted, the transmitted signal, ${{\bf{x}}_i}$, of the $i$-th user is derived by shifting the $i$-th FMW, ${{\bf{c}}_i}$, cyclically to the left by  $\tau_i \in \left\{ {0, 1, \cdots, M-1} \right\}$ places,
\begin{equation}
\label{TXsignal}
{\bf{x}}_i  =  \left\{ {x_i(0), x_i(1), x_i(2), \cdots, x_i(LN - 1)} \right\}  = \langle {\bf{c}}_i \rangle_{\tau_i}.
\end{equation}
When TDCS transmitted signals from different users pass through different single-path fading channels, the received signal for the $i$-th user, ${\mathbf{r}_i}=\left\{r_i(0), r_i(1), \cdots, r_i(LN-1) \right\}$, is expressed as
\begin{equation}
\label{RXsignal}
{\bf{r}}_i = \sum \limits_{j=1}^{U} \gamma_{i,j} {\bf{x}}_j +\mathbf{n}_i  = \sum \limits_{j=1}^{U} \gamma_{i,j} \langle {\bf{c}}_j \rangle_{\tau_j}+\mathbf{n}_i,
\end{equation}
where $\gamma_{i,j}$ is the path gain between the $i$-th and $j$-th users and $\mathbf{n}_i$ is the additive white Gaussian noise (AWGN) vector. Following the receiving procedure of traditional TDCS, the received signal, ${\mathbf{r}_i}$, is periodically correlated with the local reference FMW, ${\bf{c}}_i$, to obtain
\begin{eqnarray}
\label{RXPACF1}
\varphi_{{{\bf{r}}_i},{\bf{c}}_i} \left( \tau \right) & = & \sum \limits_{j=1}^{U}  \gamma_{i,j} \varphi_{\langle {\bf{c}}_j \rangle_{\tau_j},{\bf{c}}_i} \left( \tau \right) + \varphi_{{\mathbf{n}}_i,{\bf{c}}_i} \left( \tau \right) \nonumber\\
& = & \underbrace{\gamma_{i,i} \varphi_{\langle {\bf{c}}_i \rangle_{\tau_i},{\bf{c}}_i} \left( \tau \right)}_{ACF} \nonumber\\
& + & \underbrace{\sum \limits_{j=1,j \neq i}^{U} \gamma_{i,j} \varphi_{\langle {\bf{c}}_j \rangle_{\tau_j},{\bf{c}}_i} \left( \tau \right)}_{CCF} \nonumber \\
& + & \underbrace{\varphi_{{\mathbf{n}}_i,{\bf{c}}_i} \left( \tau \right)}_{Noise}.
\end{eqnarray}
The first item represents the periodic ACF of the $i$-th user's FMW with the effects of the cyclic shift $\tau_i$ and the path gain $\gamma_{i,i}$. Note that there is a desired periodic autocorrelation peak at $\tau = \tau_i$ corresponding to transmitted data symbols. The second item represents the periodic CCF between the $i$-th user's local reference FMW, ${\bf{c}}_i$, and the received signals from other users, $\left\{{\bf{x}}_j, j\neq i \right\}$. Generally speaking, the second item means the MUI of the $i$-th user. The third item represents the correlation output of AWGN vector.

From the correlation property of the designed sequences in \eqref{PCCFFMW}, the interference from the $j$-th user can be rewritten as follows,
\begin{flalign}
\label{PCCFFMWtau}
\gamma_{i,j} \varphi_{\langle {\bf{c}}_j \rangle_{\tau_j},{\bf{c}}_i} \left( \tau  \right) = \left\{ {\begin{array}{*{20}{l}}
\gamma_{i,j} {L  {\psi _{{b_j},{b_i}}}\left( \tau-\tau_j  \right)}, & -N + \tau_j < \tau  < N + \tau_j \\
{0,}& \text{otherwise.}
\end{array}} \right..
\end{flalign}
For the $i$-th user, the interference from the $j$-th user can be eliminated if and only if
\begin{equation}
\label{ZCZRuleAWGN}
\tau_i \not \in \left( -N + \tau_j, N + \tau_j \right).
\end{equation}
By generalizing \eqref{ZCZRuleAWGN}, the TDCS-based CRN with MUI avoidance can be achieved if and only if,
\begin{eqnarray}\label{MUIfreeAWGN}
\tau_i \not \in & \bigcup \limits_{j = 1, i \ne j}^{U} \left( -N+\tau_j, N+\tau_j \right),
\end{eqnarray}
for $i = 1, 2, ..., U$.

Recalling from the CCSK modulation scheme, the maximal shift range should be $[0,LN-1]$ for the FMW of length $LN$. However, the shift range of MUI-free TDCS-based CRNs must satisfy the constraints defined in \eqref{MUIfreeAWGN}, although the designed FMWs are also of length $LN$. In this paper, the specially-designed CCSK over a limited shift range is termed as partial cyclic code shift keying (P-CCSK).

\subsection{Design of modulation schemes in multipath fading channels}
\begin{figure}
\centering
  \includegraphics[width=4.5in]{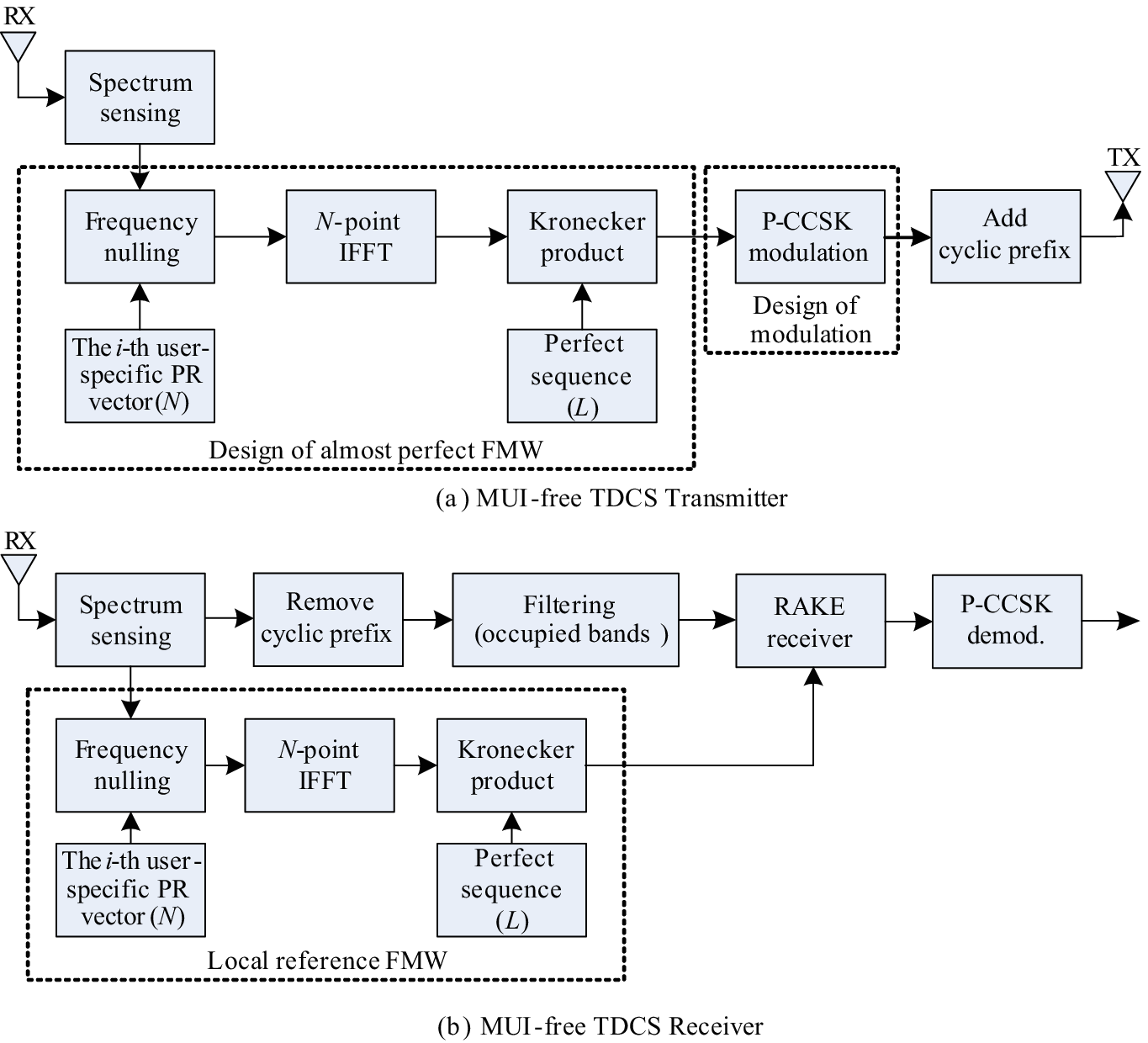}
  \caption{Block diagram of each user in MUI-free TDCS-based CRNs}
  \label{FIG:BLKTX}
\end{figure}

Now we consider the design of modulation scheme in multipath fading channels, of which the channel response can be modeled as a discrete-time finite impulse response denoted as $h_p$, where $p \in \left\{0,1, \cdots, T_{max} \right\}$, and $T_{max}$ is the maximal channel order. Note here that the indexing for calculating periodic correlation functions is based on the modulo operation on the sequence length. This results in the circular operations that can be achieved in practice by introducing cyclic prefix of appropriate length commonly seen in OFDM systems. For the TDCS transmitter shown in Fig.~\ref{FIG:BLKTX}, a cyclic prefix, which is a repetition of the last $T_g \geq T_{max}$ samples in a data block, is inserted at the beginning of each transmission data block.

We assume that all transmitted signals in multipath fading channels are same as that in single-path fading channels given in \eqref{TXsignal}. When passing through multipath fading channels, the received signal of the $i$-th user after removing cyclic prefix is given by
\begin{equation}
\label{RXsignal}
{\bf{r}}_i = \sum \limits_{j=1}^{U} \sum \limits_{p=0}^{T_{max}} h_{p}^{i, j} \langle {\bf{x}}_j \rangle_{p} +\mathbf{n}_i=\sum \limits_{j=1}^{U} \sum \limits_{p=0}^{T_{max}} h_{p}^{i, j} \langle {\bf{c}}_j \rangle_{\tau_j+p} +\mathbf{n}_i,
\end{equation}
where $h_{p}^{i, j}$ is the $p$-th path gain between the $i$-th and $j$-th users. The received signal, ${\mathbf{r}_i}$, is periodically correlated with the local reference FMW, ${\bf{c}}_i$, to obtain
\begin{eqnarray}
\label{RXPACF2}
\varphi_{{{\bf{r}}_i},{\bf{c}}_i} \left( \tau \right) & = & \sum \limits_{j=1}^{U}  \sum \limits_{p=0}^{T_{max}} h_{p}^{i, j} \varphi_{\langle {\bf{c}}_j \rangle_{\tau_j+p},{\bf{c}}_i} \left( \tau \right) + \varphi_{{\mathbf{n}}_i,{\bf{c}}_i} \left( \tau \right) \nonumber\\
& = & \underbrace{\sum \limits_{p=0}^{T_{max}} h_{p}^{i, i} \varphi_{\langle {\bf{c}}_i \rangle_{\tau_i+p},{\bf{c}}_i} \left( \tau \right)}_{ACF} \nonumber\\
& + & \underbrace{\sum \limits_{j=1, i \neq j}^{U}  \sum \limits_{p=0}^{T_{max}} h_{p}^{i, j} \varphi_{\langle {\bf{c}}_j \rangle_{\tau_j+p},{\bf{c}}_i} \left( \tau \right)}_{CCF} \nonumber\\
& + & \underbrace{\varphi_{{\mathbf{n}}_i,{\bf{c}}_i} \left( \tau \right)}_{Noise}.
\end{eqnarray}
The first item represents the periodic ACF of the $i$-th user's FMW with the effects of multipath fading channels, while the second item represents the MUI from other users also with the effects of multipath fading channels. Focusing on the first item, there are a few periodic autocorrelation peaks in the range of $[\tau_i, \tau_i+T_{max}]$ corresponding to each path of the multipath fading channel. Similarly with the scenario of single-path fading channels discussed above, the $i$-th user's receiver can avoid the interference from the $j$-th user if and only if
\begin{equation}
\varphi_{\langle {\bf{c}}_j \rangle_{\tau_j+p},{\bf{c}}_i} \left( \tau \right)=0, p \in [0, T_{max}], \tau \in [\tau_i, \tau_i+T_{max}].
\end{equation}
Hence, we have
\begin{equation}
\label{ZCZRule}
\tau_i \not \in \left( -N + \tau_j-T_{max}, N + \tau_j+T_{max} \right).
\end{equation}
By generalizing \eqref{ZCZRule}, the TDCS-based CRN with MUI avoidance in multipath fading channels can be achieved if and only if,
\begin{eqnarray}\label{MUIfreeCondArray}
\tau_i \not \in & \bigcup \limits_{j = 1, i \ne j}^{U} \left( -N+\tau_j-T_{max}, N+\tau_j+T_{max} \right),
\end{eqnarray}
for $i = 1, 2, ..., U$. By following the constraints in \eqref{MUIfreeCondArray}, \eqref{RXPACF2} can be rewritten as
\begin{equation}
\label{RXPACF3}
\varphi_{{{\bf{r}}_i},{\bf{c}}_i} \left( \tau_i \right) = \underbrace{\sum \limits_{p=0}^{T_{max}} h_{p}^{i, i} \varphi_{\langle {\bf{c}}_i \rangle_{\tau_i+p},{\bf{c}}_i} \left( \tau_i \right)}_{ACF} + \underbrace{\varphi_{{\mathbf{n}}_i,{\bf{c}}_i} \left( \tau_i \right)}_{Noise}.
\end{equation}

Note that the scheme provides MUI-free transmission whenever the maximum delay spread of the channel does
not exceed the specified limit. Meanwhile, it converts the multiple access problem over a multipath fading channel to a traditional single-user CDMA problem over a multipath fading channel. As shown in Fig.~\ref{FIG:BLKTX}, the maximal-ratio combining (MRC)-RAKE receiver can be employed here to exploit channel diversity.

\section{Further Discussions}
\subsection{Multiuser capability}
Multiuser capability plays an important role when measuring a specific network. For the sake of simplicity in this paper, only the multiuser capability in single-path fading channels is derived by following the constraints in \eqref{MUIfreeAWGN}, and that in multipath fading channels can be derived in the same way under the constraints in \eqref{MUIfreeCondArray}.

\begin{figure}
\centering
  \includegraphics[width=4.5in]{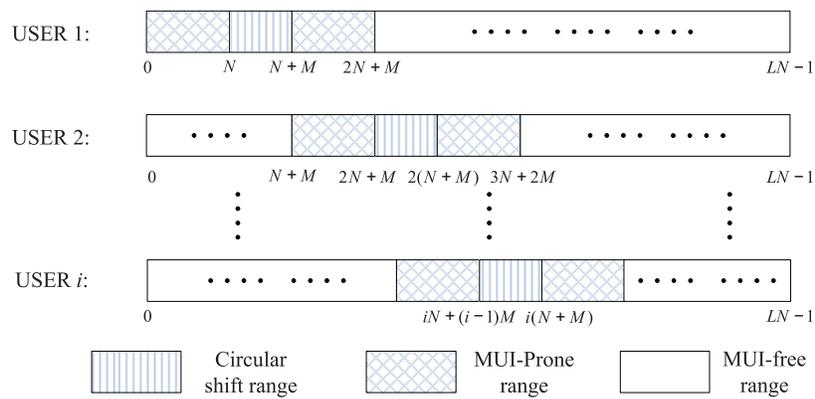}
  \caption{Example of MUI-free CRNs with $M$-ary P-CCSK to satisfy the constraints in \eqref{MUIfreeAWGN}}
  \label{FIG:MUIFree}
\end{figure}

Fig.~\ref{FIG:MUIFree} shows an example of the transmitted signals of MUI-free TDCS-based CRNs in which each user adopts $M$-ary P-CCSK. For the first user, when the circular shift range is $\tau_1 \in [N, N+M-1]$, the MUI-prone region is $[0,2N+M-1]$. From the constraints in \eqref{MUIfreeAWGN}, the second user chooses the neighboring $M$ positions starting from $2N+M$, and the circular shift region is $\tau_2 \in [2N+M, 2N+2M-1]$. Similarly, the circular shift range for the $i$-th user is $\tau_i \in [iN+(i-1)M, i(N+M)-1]$. As a result, the maximal possible number of users, $U_{Max}$, is derived by
\begin{equation}\label{NUMUSER}
U_{Max} = \lfloor \frac{LN}{N+M} \rfloor,
\end{equation}
where $\left\lfloor {p/N} \right \rfloor $  denotes the largest integer smaller than $p/N$. Table I lists the multiuser capability of different parameter configurations. For any given $L$, a smaller value of $M/N$ results in a larger number of users. Meanwhile, the multiuser capability improves as $L$ increases.

\begin{table}[ht]
\caption{Multiuser capability of MUI-free TDCS-based CRNs}
\centering
\begin{tabular}{|c|c|c|c| }
\hline
  & $M/N= 1/4$ & $M/N = 1$ & $M/N = 2$ \\ \hline\hline
  $L=8$ & 6 & 4 & 2 \\
  \hline
  $L=9$ & 7 & 4 & 3 \\
  \hline
  $L=12$ & 9 & 6 & 4 \\
  \hline
  $L=16$ & 12 & 8 & 5 \\
  \hline
\end{tabular}
\end{table}

\subsection{Aggregated throughput}
Now we discuss the aggregated throughput of MUI-free TDCS-based CRNs. Since the order of P-CCSK must be a power of 2, i.e., $2^k$ for $ k=1, 2, \cdots$, the maximal possible order of P-CCSK for a given number of users, $U$, is derived from \eqref{NUMUSER} as follows,
\begin{equation}
\label{Modorder}
M_{max} = 2^ {\lfloor \log_2(\frac{LN}{U}-N) \rfloor},
\end{equation}
From \cite{Fumat}, since the spectrum efficiency of each user is given by
\begin{equation}\label{EffTr}
\eta_{max} = \frac {\log _2 (M_{max})}{\beta L N} \text{(bps/Hz)},
\end{equation}
where $\beta$ denotes the portion of available spectrum bins, the achievable aggregated throughput can be expressed as
\begin{equation}\label{EffAg}
\eta _{Agg} =  U \eta_{max} = \frac {U {\lfloor \log_2(\frac{LN}{U}-N) \rfloor} }{\beta L N} \text{(bps/Hz)}.
\end{equation}

\begin{figure}
\centering
  \includegraphics[width=4.5in]{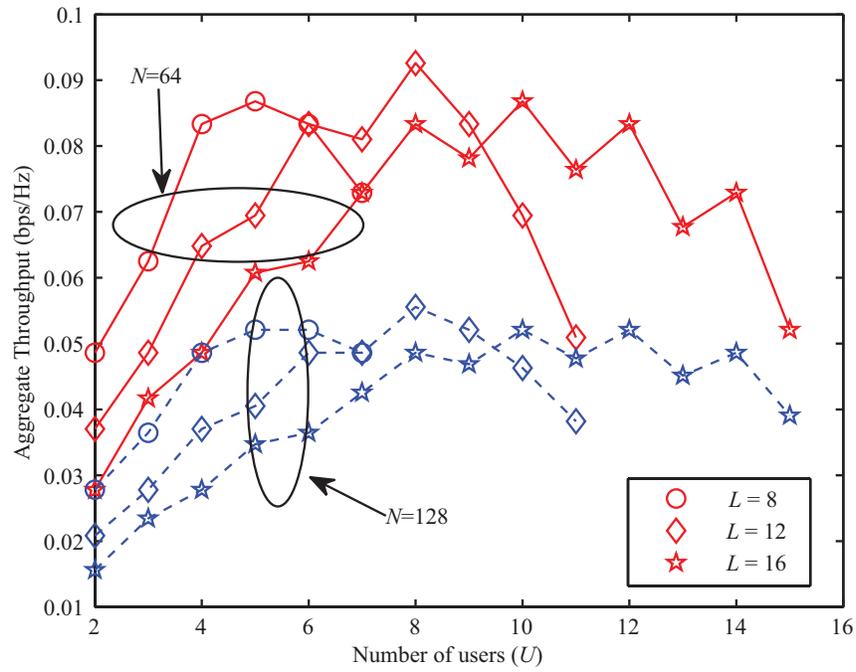}
  \caption{Aggregated throughput of MUI-free TDCS-based CRNs ($\beta=3/4$)}
  \label{FIG:Aggth}
\end{figure}

According to the aggregated throughput shown in Fig.~\ref{FIG:Aggth}, the following remarks are relevant.
\begin{itemize}
  \item Increasing the number of spectrum bins, $N$, results in the lower aggregated throughput. Fig.~\ref{FIG:Aggth} shows two groups of curves for  $N=64$ and 128, respectively. With the fixed values of $U$ and $L$ in \eqref{EffAg}, it is easy to prove that $\eta _{Agg} \left( N; U, L \right)$ is a decreasing function of $N$.
  \item When the length of FMWs (spreading factor) is fixed, such as $\left\{L=8,N=128\right\}$ and $\left\{L=16,N=64\right\}$, the maximal possible number of users and the aggregated throughput supported by MUI-free CRNs increase as $N$ decreases.
\end{itemize}
Consequently, the above observations provide a rule of thumb for MUI-free TDCS-based CRNs to satisfy the requirement of desired number of users and aggregated throughput.

\begin{table*}
\newcommand{\tabincell}[2]{\begin{tabular}{@{}#1@{}}#2\end{tabular}}
\label{table1}
\centering
\caption{System parameters for different random-spreading-based CRNs}
\begin{tabular}{|c|c|c|c|c|}
  \hline
   Parameters & MC-CDMA \cite{Hara} & Trad. TDCS \cite{Swackhammer} & MUI-free TDCS \\
   \hline
   \hline
  Spreading codes & Zadoff-Chu & Pseudorandom & Almost perfect \\
  \hline
   Codes of length & 1024 & 1024 & 1024 ($L=16, N=64$) \\
  \hline
  Orthogonality & No & No & Yes \\
   \hline
  Modulation & QPSK & CCSK & P-CCSK \\
  \hline
  Receiver & MMSE-FDE & MMSE-FDE & Rake Receiver \\
  \hline
  Channels & \multicolumn{3}{|c|}{Single-path and Multipath fading channels} \\
  \hline
  Spectrum utilization pattern & \multicolumn{3}{|c|}{Bandwidth: 10MHz, N/A bands: 2.5$\sim$3.75 MHz and 6.25$\sim$7.5 MHz}\\
  \hline
  \end{tabular}
\end{table*}

\section{Simulation Results}
Performances of different random-spreading-based CRNs, i.e., MC-CDMA, traditional TDCS and MUI-free TDCS, are demonstrated via simulations, and system parameters are shown in Table II. In order to make a fair comparison, Zadoff-Chu sequences are used for MC-CDMA systems since they produce significantly better average bit error probability than Walsh and Gold sequences \cite{PoPo}. For MUI-free TDCS-based CRNs, the steps to generate a class of almost perfect  sequences are: 1) generating a class of basis FMWs according to the method introduced in \cite{Swackhammer}, where each one is of length $N=64$, and 2) generating a class of almost perfect sequences according to \eqref{Krongen}, where the quadriphase perfect sequence of length $L=16$ is
\begin{eqnarray}
\mathbf{A} = \left\{1, 1, 1, 1, 1, j, -1, -j,
 1, -1, 1, -1, 1, -j, -1, j \right\}  \nonumber.
\end{eqnarray}

\subsection{BER performance in single-path fading channels}
\begin{figure}
\centering
  \includegraphics[width=4.5in]{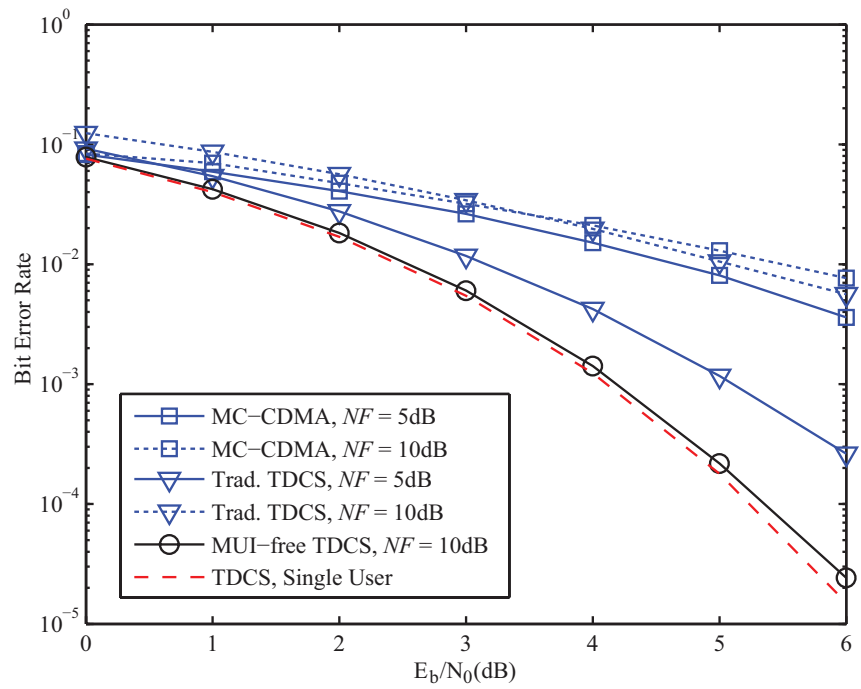}
  \caption{BER performance of different CRNs in single-path fading channels ($U=4$)}
  \label{FIG:BERCompNonFullloadedU4}
\end{figure}

Fig.~\ref{FIG:BERCompNonFullloadedU4} shows BER performance of different CRNs $(U=4)$. The MUI-free TDCS is partial loading as the order of P-CCSK, $M=N$, is less than $M_{max}$ defined in \eqref{Modorder}. Owing to the non-orthogonality of spreading sequences, the BER performance of MC-CDMA and traditional TDCS is rapidly degraded when the near-far factor, $NF$, increases. In contrast, the MUI-free TDCS is robust against the near-far problem. When $U= 4$ and $NF=10$dB, the simulated BER performance is nearly the same as that of the single-user case, which validates the capability of MUI avoidance.

\begin{figure}
\centering
  \includegraphics[width=4.5in]{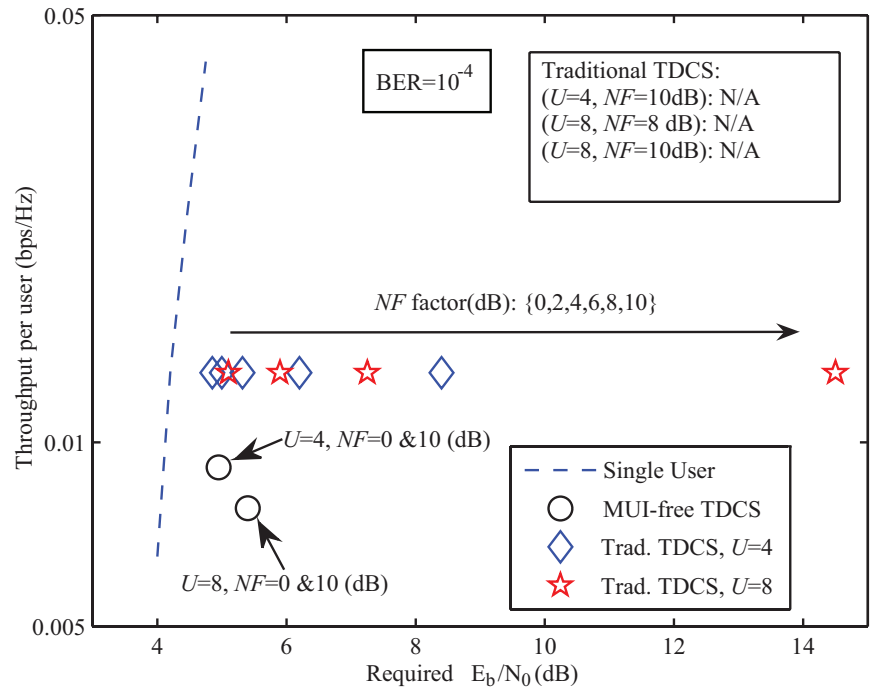}
  \caption{Throughput per user versus the required $E_b/N_0$ to achieve BER=$10^{-4}$}
  \label{FIG:BERCompFullloaded}
\end{figure}

Fig.~\ref{FIG:BERCompFullloaded} shows the throughput per user versus the required $E_b/N_0$ to achieve BER=$10^{-4}$. In case of full loading, each user of MUI-free TDCSs chooses the maximal possible modulation order to maximize spectrum efficiency. For the traditional TDCS, the MUI dominates BER performance and results in an error floor as the number of users, $U$, and the near-far factor, $NF$, increase, e.g., the BER performance of traditional TDCSs cannot achieve $10^{-4}$ when $U=4, NF=10$dB and $U=8, NF=\left\{8, 10 \right\}$dB. Meanwhile, it is interesting to note that, there is a performance gap compared to the single-user case with the full loading. If $M_{max}$-ary P-CCSK is adopted, the limited shift range allows each transmitted signal of length $LN$ carrying $\log_2(M_{max})$ bits, which is less than $\log_2(LN)$ bits in the single-user case. This performance loss is not avoidable. For example, the MUI-free TDCS with $U=4$ and $U=8$ suffer from $0.85$dB and $1.35$dB performance loss in terms of $E_b/N_0$.

\subsection{BER performance versus near-far factors}
\begin{figure}
\centering
  \includegraphics[width=4.5in]{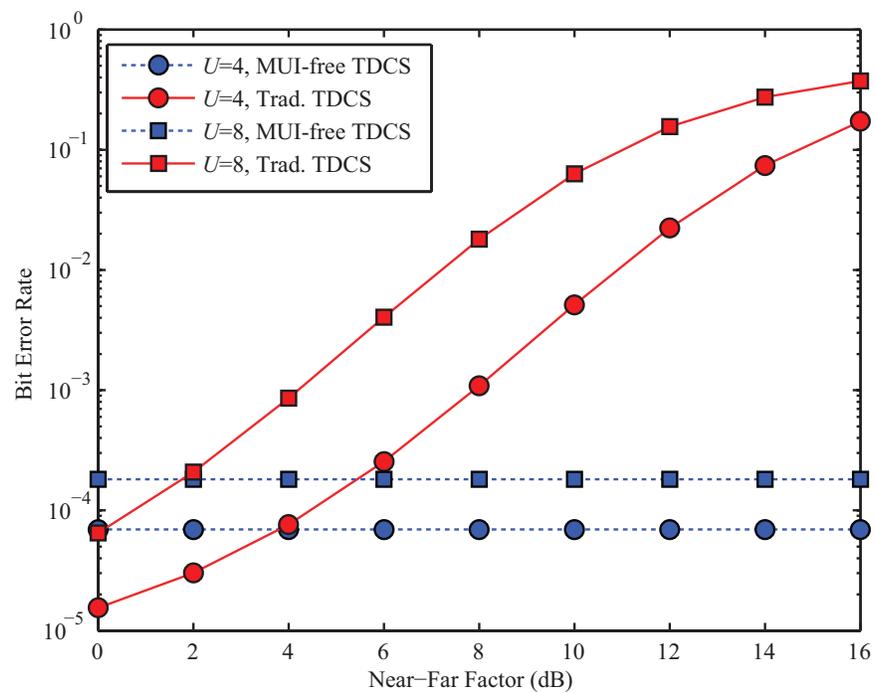}
  \caption{BER performance versus near-far factors for $E_b/N_0 = 5$dB, $U=\left\{4,8\right\}$}
  \label{FIG:BERComp}
\end{figure}

Fig.~\ref{FIG:BERComp} shows the effects of near-far factors and number of users on BER performance. In this simulation, we consider the number of users, $U=\left\{4, 8\right\}$, and the near-far factors, $NF = \left\{0, 2, 4, \cdots, 16\right\}$dB. As $U$ and $NF$ increase, BER performance of traditional TDCSs degrades rapidly, while that of MUI-free TDCS is hardly affected. For practical decentralized CRNs, such as CR Ad-hoc wireless sensing network with significant near-far effect, the MUI-free TDCS can be considered as a preferable candidate when power control techniques cannot be used.

\subsection{Spectrum sensing mismatch}
According to IEEE 802.22 \cite{Stevenson}, channel uncertainties are likely to produce a mismatch between the spectrum bins identified by the spectrum sensing results obtained by the transmitter and receiver. To discuss the performance with imperfect spectrum sensing, we use the correlation coefficient $\eta$ to quantify the mismatch, as shown in \cite{Martin}\cite{LOWPAPR}.

\begin{figure}
\centering
  \includegraphics[width=4.5in]{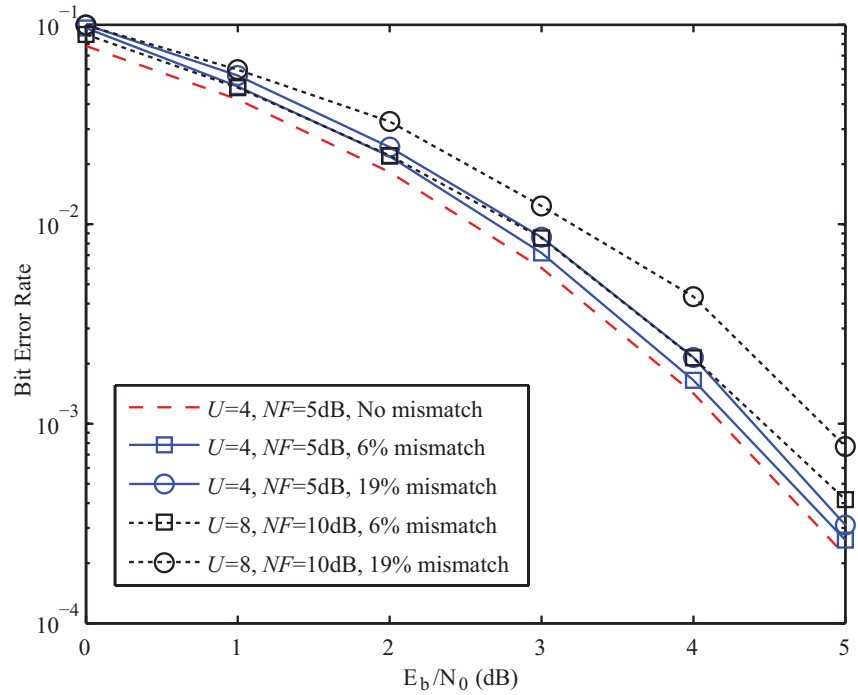}
  \caption{BER performance of MUI-free TDCS-based CRNs in case of spectrum sensing mismatch}
  \label{FIG:BERMISMATCH}
\end{figure}

Fig.~\ref{FIG:BERMISMATCH} shows the BER performance of MUI-free TDCS with spectrum sensing mismatch. With a small number of users, i.e., $U=4$, each user achieves a BER performance similar to that with perfect spectrum sensing, with only $0.15$dB degradation in terms of $E_b/N_0$ at BER = $10^{-3}$. However, with a larger number of users, i.e., $U=8$, the BER performance is obviously worse than that with perfect spectrum sensing. This is because the constraints in \eqref{MUIfreeAWGN} for MUI avoidance are no longer satisfied in the spectrum sensing mismatch scenario, and the residual MUI degrades the BER performance.

\subsection{BER performance in multipath fading channels}
\begin{figure}
\centering
  \includegraphics[width=4.5in]{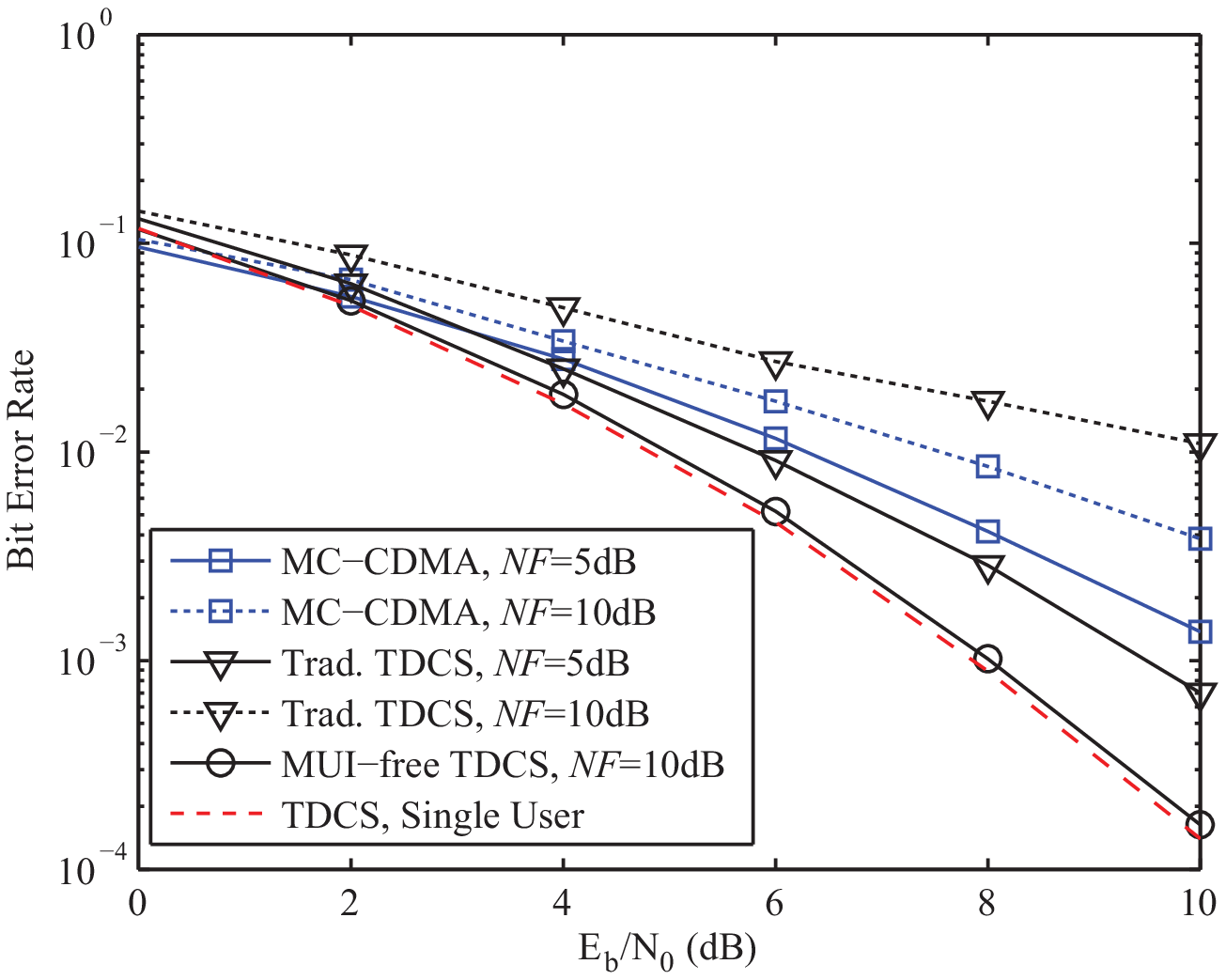}
  \caption{BER performance of different CRNs in multipath fading channels ($U=4$)}
  \label{FIG:BERMultipath}
\end{figure}

Now we discuss BER performance of different random-spreading-based CRNs in multipath fading channels such as the COST207RAx6 channel \cite{Cost207}. In this simulation, the length $1/4$ cyclic prefix is adopted to eliminate the effects of multipath fading channels, and minimal mean square error frequency-domain equalization (MMSE-FDE) is considered for MC-CDMA and traditional TDCS. As shown in Fig.~\ref{FIG:BERMultipath}, MC-CDMA and traditional TDCS still suffer the MUI due to the non-zero periodic CCF between any pair of transmitted signals, while the MUI-free TDCS achieves the same BER performance as that of the single-user case. In addition, the MRC-RAKE receiver can effectively exploit the channel diversity of multipath fading channels. This simulation verifies that, even in case of multipath fading channels, the MUI-free TDCS is also an effective solution against the near-far problem and multiuser interference.

\section{Conclusion}
In this paper, a novel framework based on the joint designs of sequences and modulation schemes is presented to realize the MUI-free TDCS-based CRNs. It is shown that, when the spreading sequence of length $N$ in the frequency domain and that of length $L$ in the time domain are carefully chosen, a class of almost perfect sequences is obtained through two-dimensional time-frequency synthesis. That is, periodic ACF or CCF is identically zero for $\left(L-2 \right)N+1$ different values of relative circular shift. Then, the specially-designed CCSK over a limited shift range is introduced to achieve the advantage of MUI avoidance in both of single-path and multipath fading channels. By analyzing the multiuser capability and aggregated throughput, we provide a rule of thumb for MUI-free TDCS-based CRNs with a given number of users and aggregated throughput. Simulation results demonstrate that the proposed scheme achieves significant BER performance improvements compared to these existing random-spreading-based CRNs. For practical decentralized CRNs, such as CR Ad-hoc wireless sensing network with significant near-far effect, the MUI-free TDCS-based CRN is considered as a preferable candidate when power control techniques cannot be used.

\newpage

\newpage

\newpage

\newpage

\newpage


\newpage

\newpage

\newpage

\newpage


\begin{thebibliography}{l}

\bibitem{Haykin}
S. Haykin, "Cognitive radio: brain-empowered wireless communications," \emph{IEEE J.  Select. Areas Commun.}, vol.23, no.2, pp.201-220, 2005

\bibitem{Boroujeny}
B.F. Boroujeny and R. Kempter, "Multicarrier communications techniques for spectrum sensing and communications in cognitive radio," \emph{IEEE Commun. Mag.}, vol.46, pp.80-85, 2008

\bibitem{RobertSP}
M. Roberts, M.A. Temple, R.A. Raines, R.F. Mills, and M.E. Oxley, "Communication waveform design using an adaptive spectrally modulated, spectrally encoded (SMSE) framework," \emph{IEEE J. Sel. Areas Signal Process.}, vol.1, no.1, pp.203-213, 2007

\bibitem{Chakrava}
V. Chakravarthy, X. Li, Z. Wu, M.A. Temple, F. Garber, R. Kannan, and A. Vasilakos, "Novel overlay/underlay cognitive radio waveforms using SD-SMSE framework to enhance spectrum efficiency--part I: theoretical framework and analysis in AWGN channel," \emph{IEEE Trans. Commun.}, vol.57, no.12, pp.3794-3804, 2009

\bibitem{Hara}
S. Hara and R. Prasad, "Overview of multi-carrier CDMA," \emph{IEEE Commun. Mag.}, vol.35, no.12, pp.126-133, 1997

\bibitem{Budiarjo}
I. Budiarjo, H. Nikookar, and L.P. Ligthart, "Cognitive radio modulation techniques," \emph{IEEE
Signal Process. Mag.}, vol.25, no.6, pp.24-34, 2008

\bibitem{Fumat}
G. Fumat, P. Charge, A. Zoubir, and D.F. Prunaret, "Transform domain communication systems from a
multidimensional perspective impacts on bit error rate and spectrum efficiency," \emph{IET
Commun.}, vol.5, no.4, pp.476-483, 2011

\bibitem{QCCSK}
S. Hu, G. Bi, Y.L. Guan, and S.Q. Li, "Spectrally efficient transform domain communication system with quadrature cyclic code shift keying," \emph{IET Commun.}, vol. 7, no. 4, pp.382¨C390, 2013

\bibitem{Cluster}
S. Hu, Y.L. Guan, G. Bi, and S.Q. Li, "Cluster-based transform domain communication systems for high spectral efficiency," \emph{IET Commun.}, vol.6, no.16, pp.2734-2739, 2012

\bibitem{Chakravarthy-TDCS}
V. Chakravarthy, A.S. Nunez, and J.P. Stephens, "TDCS, OFDM, and MC-CDMA: a brief tutorial,"
\emph{IEEE Radio Commun.}, vol.43, no.9, pp.S11-S16, 2005

\bibitem{Letaief}
K.B. Letaief, W. Zhang, "Cooperative communications for cognitive radio networks," \emph{Proceedings of the IEEE}, vol.97, no.5, pp.878-893, 2009

\bibitem{Tevfik}
T. Yucck and H. Arslan, "A survey of spectrum sensing algorithms for cognitive radio applications," \emph{IEEE Comm. Survey Tutorials}, vol. 11, no.1, pp.116-130, 2009

\bibitem{PoPo}
B.M. Popovic, "Spreading sequences for multicarrier CDMA systems," \emph{IEEE Trans. Commun.}, vol.47, no.6, pp.918-926, 1999

\bibitem{Zadoff}
R.L. Frank and S.A. Zadoff, "Phase shift pulse with good periodic correlation properties," \emph{IRE Trans. Inform. Theory}, vol. IT-8, pp. 381-382, 1962

\bibitem{Chu}
D.C. Zhu, "Polyphase codes with good periodic correlation properties," \emph{IEEE Trans. Inform. Theory}, vol. IT-18, pp. 531-533, 1972

\bibitem{Feng}
S. Feng, H. Zheng, H. Wang, J. Liu and P. Zhang, "Preamble design for non-contiguous spectrum usage in cognitive radio networks," in \emph{Proc. IEEE Wireless Commun. and Networking Conf. (WCNC 2009)}, pp.1-6, 2009

\bibitem{Swackhammer}
P.J. Swackhammer, M.A. Temple and R.A. Raines, "Performance simulation of a transform domain
communication system for multiple access applications," in \emph{Proc. IEEE Military Commun.
Conf.}, vol.2, pp.1055-1059, 1999




\bibitem{Dillard}
G.M. Dillard, M. Reuter, J. Zeidler, and B. Zeidler, "Cyclic code shift keying: a low probability
of intercept communication technique," \emph{IEEE Trans. Aerosp. Electron. Syst.}, vol.39, no.3,
pp.786-798, 2003

\bibitem{Stark}
W.E. Stark and D.V. Sarwate, "Kronecker sequences for spread-spectrum communication," \emph{IEE Proc.}, vol.128, no.2, pp.104-109, 1981

\bibitem{Stevenson}
C.R. Stevenson, G. Chouinard, Z. Lei, W. Hu, S.J. Shellhammer and W. Caldwell, "IEEE 802.22: the
first cognitive radio wireless regional area network standard," \emph{IEEE Commun. Mag.}, vol. 47,
no. 1, pp. 130-138, 2009.

\bibitem{Martin}
R.K. Martin and M. Haker, "Reduction of peak-to-average power ratio in transform domain
communication systems", \emph{IEEE Trans. Wireless Commun.}, vol. 8, no. 9, pp. 4400-4405, 2009.

\bibitem{LOWPAPR}
S. Hu, G. Wu, Y. Xiao, X. Lei, and S.Q. Li, "Design of Low PAPR Fundamental Modulation Waveform for Transform Domain Communication System," \emph{Wireless Personal Commun.}, vol. 71, no. 3, pp.2215-2229, 2013

\bibitem{Cost207} 3GPP Specification Series: "Radio transmission and reception, 3GPP TS
    45.005", 2007.

\end{thebibliography}
\end{document}